\documentclass[reprint,aps,superscriptaddress]{revtex4-2}
\usepackage{graphicx}
\graphicspath{{figures/}}
\usepackage{dcolumn}
\usepackage{braket}
\usepackage{bm}
\usepackage{units}
\usepackage{notes2bib}

\bibnotesetup{
note-name = ,
use-sort-key = false
}

\begin{document}
\preprint{APS/123-QED}
\title{Universal method to extract the average electron spin relaxation in organic semiconductors from muonium ALC resonances}
\author{Licheng Zhang}
\address{School of Physical and Chemical Sciences,
Queen Mary University of London,
London, E1 4NS, UK}
\author{J. S. Lord}
 \email{James.Lord@stfc.ac.uk}
\address{ISIS Muon and Neutron Source, STFC,
Rutherford Appleton Laboratory, Harwell Campus, Didcot, OX11 0QX, UK}
\author{A. J. Drew}
 \email{A.J.Drew@qmul.ac.uk}
\address{School of Physical and Chemical Sciences,
Queen Mary University of London,
London, E1 4NS, UK}

\begin{abstract}
 Muon spin spectroscopy and in particular the avoid level crossing (ALC) technique is a sensitive probe of electron spin relaxation (eSR) in organic semiconductors. In complex ALC spectra, eSR can be challenging to extract, as it requires the modelling of overlapping ALCs, where covariance between parameters can result in significant uncertainties. Here we demonstrate a general method to extract eSR rate, which is independent on the number of ALCs resonances present, whether they overlap or not, and what the muonium hyperfine (isotropic and anisotropic) parameters are. This can then be used to extract an accurate value for eSR rate and as guidance for undertaking experiments efficiently.
\end{abstract}

\maketitle

Spin interactions are vital to many applications of organic semiconductors (OSC), such as organic spintronics \cite{naber2007organic,morley2008room}, display technologies \cite{geffroy2006organic} and photovoltaics \cite{brabec2011organic}, with important mechanisms such as the various unpaired electron spin relaxation mechanisms, inter-system crossing, singlet fission and thermally activated delayed flourescence (TADF) \cite{ForrestOrganic}. Having an accurate measurement of the electron spin relaxation time, $\tau_e$, that can be applied to a wide variety of molecules is vital for developing these technologies, but also an interest in its own right. Experimental measurements of $\tau_e$ in OSC vary between a less than a microsecond to potentially second or more \cite{cinchetti2009determination,dediu2002room,schott2017tuning}.

Electron paramagnetic resonance (EPR) is  capable of providing a measure of $\tau_e$, but it requires a material with intrinsic unpaired electrons or electrons provided by doping, which is not always possible. $\tau_e$ can also be extracted from magnetotransport measurements on unipolar spin valve devices \cite{dediu2009spin,pramanik2007observation},
but this approach requires a robust theoretical model. For example,  models that build on the Elliott-Yafet \cite{elliott1954theory}
or Dyakonov-Perel mechanisms \cite{d1971spin},
or one involving electrons observing a random hyperfine field upon hopping between molecular sites \cite{bobbert2009theory}.
There are a number of issues with many of the earlier theoretical models. They are usually most applicable to band semiconductors, which are likely not suited to the localised molecular orbitals in OSC. All of the theoretical models that are transport based that require an accurate measure of the charge carrier mobility, whose uncertainty even in well-defined organic light emitting diodes (OLED) leads to a significant uncertainties in the value of $\tau_e$. The strong temperature dependence of charge carrier mobility also often obscures the underlying physics of the spin relaxation mechanisms. 

Harmon et al. have produced a theory for spin relaxation of hopping carriers in a disordered organic semiconductor, that considers hopping induced eSR from spin-orbit coupling (SOC) and the hyperfine interaction (HFI) and an ``intrasite'', or intramolecular, spin relaxation mechanism involving spin-lattice interactions (time varying HFI) but not SOC \cite{Harmon2013}. Charge carrier mobility in OSC, especially in devices where the OSC is amorphous, is low; the charge carrier resides on the molecule for a significant amount of time, and it is not clear whether the hopping or intramolecular mechanisms are dominant under what limits in real-world OSC. Experimental verification is needed. More recently, a particularly elegant systematic but primarily theoretical study has been carried out that demonstrates a rich variability of the g-shifts with the effective SOC \cite{schott2017tuning}, which is dependant on subtle aspects of molecular composition and structure. The work was carried out in isolated molecules and the authors note their results are likely to be relevant in solid state systems, although this is not 100\% clear and would also benefit from being experimentally determined. In general, for intramolecular interactions, the relative importance of SOC or HFI has not been properly addressed experimentally. 
 
 Muon spin resonance ($\mu$SR), also known as avoided level crossing (ALC) spectroscopy, has been demonstrated to be an excellent probe of spin dynamics in OSC \cite{schulz2011importance, nuccio2011electron, nuccio2013importance, nuccio2014muon, han2015muonium} in the solid state and liquid/solution state, but to extract eSR requires modelling and for complex signals, this is not always possible\bibnote{We note that we are exclusively considering $\Delta_1$ resonances present in the solid state due to anisotropy, without the additional proton-electron coupling necessary for $\Delta_0$. Further details of different ALC states can be found in \cite{nuccio2014muon}.}. Importantly, it measures the intramolecular eSR, and not one based on hopping, and so is an ideal technique to study the intrinsic interactions responsible for eSR. In a muon experiment, a 100\% spin polarised positively charged muon beam is implanted into the sample. In OSC (and certain other materials), the muon can capture an electron to form a muonium atom. It loses some of its polarisation as a result of HFI between the muon and electron that result in Rabi oscillations. Muonium is often referred to as light hydrogen, by virtue of its ionisation energy and chemical reactions being similar to hydrogen, but is around 9 times lighter \cite{walker1983muon}.  In most OSC, muonium chemically bonds to a molecule (often referred to as a “muoniated radical” \cite{blundell2004muon}), where the muon-electron HFI is smaller, due to the spreading out of the electron's wavefunction on the molecule. It is worth noting that in this case, the technique self dopes an electron to the molecule, and so is applicable to the majority of OSC, unlike EPR. A relaxation of this electron's spin can induce changes in the muon spin, which are then subsequently detected \cite{nuccio2014muon}. The muon and muonium have a half-life of 2.2 $\mu$s, decaying into a positron that is correlated to the muon's spin at the time of decay. By measuring the time dependence of the spatial  distribution of emitted positrons, one is able to extract a direct measurement of the spin polarization of the ensemble of muons, and thus a measure of the electron spin dynamics in the sample. 
\begin{figure}[!h]
\centering
\includegraphics[width=\linewidth]{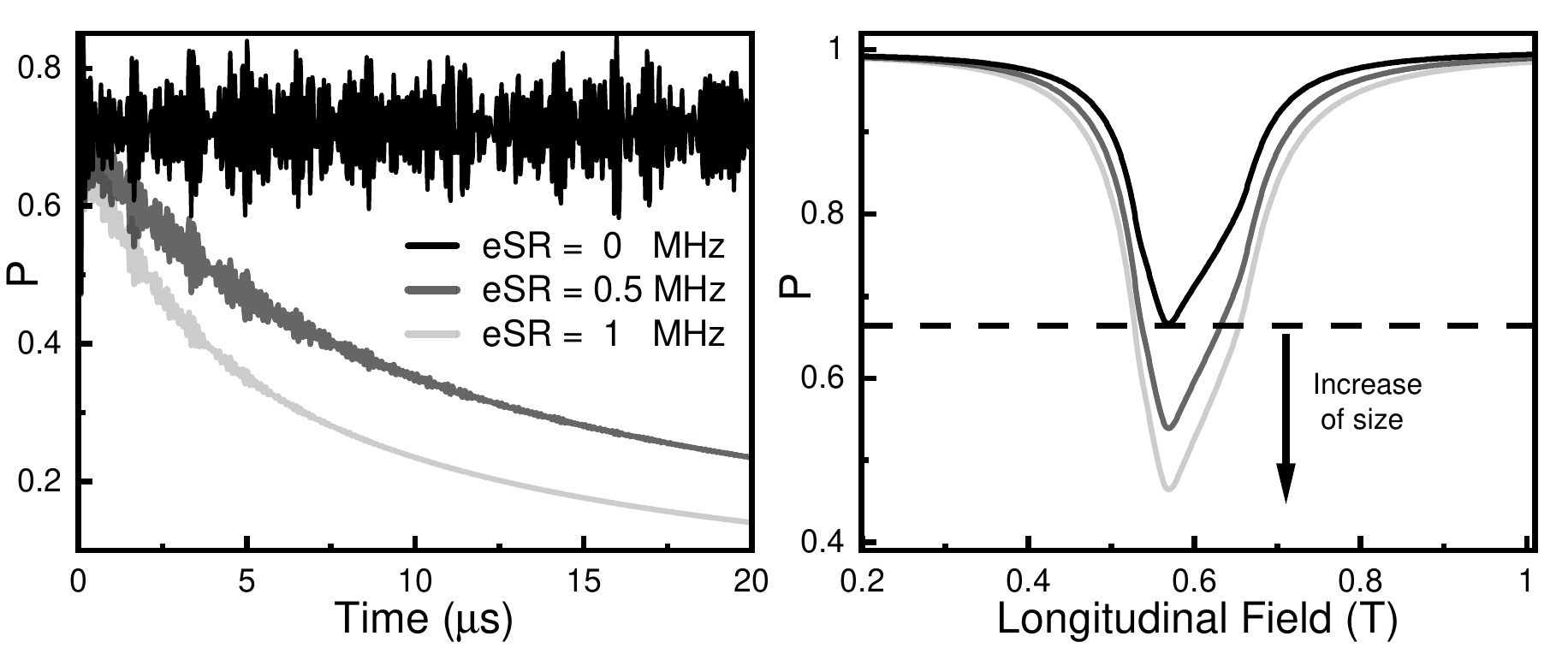}
\caption{\label{F_ALC2}The influence of the dynamical effect of the electron near an ALC(HFCC A = 160MHz, D = 20MHz, E = 10MHz). (a) The eSR effects a reduction of the muon spin polarization over time. B = 0.6T. (b) Time integral of the muon spin polarization curves.}
\end{figure}

In a typical solid-state experiment in a polycrystalline OSC, one applies a magnetic field parallel with the muon's initial spin direction, termed a longitudinal field (LF). The energy levels of the singlet/triplet bound muonium states are tuned via the Zeeman interaction. This has two main effects. Firstly it slowly decouples the muon's spin from the electron's spin as the externally applied field scales with the internal muon-electron HFI (randomly oriented with respect to the LF), known as repolarisation. Secondly, in high magnetic fields, where most attempts to measure $\tau_e$ are carried out, the eigenstates of the spin systems are to a good approximation pure Zeeman product states. Muons injected with their spins parallel or anti-parallel to the field are thus in an eigenstate, and no time evolution of spin polarization is expected. At a particular longitudinal field cross relaxation effects produce an avoided level crossing (ALC), at what would otherwise be energy level degeneracies; the m$_s$=1 and m$_s$=0 triplet eigenstates are mixtures between two Zeeman states. This leads to an oscillation between the levels. Whenever the two mixing levels belong to different muon magnetic quantum states, the time evolution causes a depolarization of the muon's spin. The muon technique is particularly sensitive to eSR whilst on an ALC resonance, as  this loss of spin polarisation is ``locked in'' as the m$_s$=-1 triplet state is off-resonance \cite{nuccio2014muon,blundell2004muon}.

In the solid state, the positions and linewidths of the ALC resonances are determined by the muon-electron isotropic and anisotropic HFI, which means that different muonium bonding sites on the molecules will have different ALC resonances. Further details can be found elsewhere \cite{nuccio2014muon, blundell2004muon, wang2016spintronic}. It has been demonstrated that eSR probed with the muonium electron via ALC spectroscopy are governed by the same mechanisms in OLEDs and optical spectroscopy \cite{nuccio2013importance}, opening up a world of opportunity to study OSC and suggesting that these effects are dominated by intramolecular mechanisms.

eSR is strongly temperature dependent because it is driven by vibrations. Typically when performing an experiment, one would measure an ALC spectra at a low temperature (e.g 10K), followed by the higher temperatures at which eSR is to be measured. One then assumes a negligibly small eSR at low temperatures to extract the eSR at high temperatures. This assumes that temperature dependent changes to muonium formation are negligible, which can be checked by transverse field measurements and by studying more than one ALC resonance (if available, and sufficiently separated in magnetic field). It is thus a good technique to extract detailed information on how SOC depends on subtle aspects of molecular composition and structure in the solid state \cite{schott2017tuning}. However, one of the main problems with using ALC $\mu$SR to probe eSR in the solid state is that often, there are overlapping ALC resonances from different muonium bonding sites, that can make fitting or modelling challenging \cite{nuccio2013importance}. It can be difficult to extract the full anisotropic technique hyperfine parameters of the different states needed to extract the eSR and there is covariance between parameters. In the extreme limit, this limitation can obscure the results entirely \cite{he2015muon}. Even in better behaved systems, there is evidence that there is a distribution of bonding angles meaning a distribution of muon-electron HFI \cite{Berlie2022, Clowney1996}. In any experiment, it can be extremely hard to justify enough time on a muon instrument to measure the complete temperature dependence with sufficient number of data points on the ALCs to extract the temperature dependent eSR. Previously, a single-point tracking of the ALC amplitude as a function of temperature has been used \cite{schulz2011importance}, but if the data involves overlapping peaks or the peak position is a function of temperature (muon-electron HFI may be a function of temperature), then this method is challenging and likely won't yield satisfactory results. 

Here, we demonstrate these limitations explicitly and demonstrate a new method to deal with some of the challenging data sets discussed above. We demonstrate that by proper handling of the data, one is able to use a universal law to extract the eSR, that is independent of the technique hyperfine parameters (and thus their temperature or bonding angle dependence) and the complexity or overlap of the ALC resonance spectra (and thus is applicable for even the extreme cases \cite{he2015muon}).  We then test it against a known and published data-set originally analysed using the historical modelling methodology.

\begin{figure*}
\centering
\includegraphics[scale=0.75]{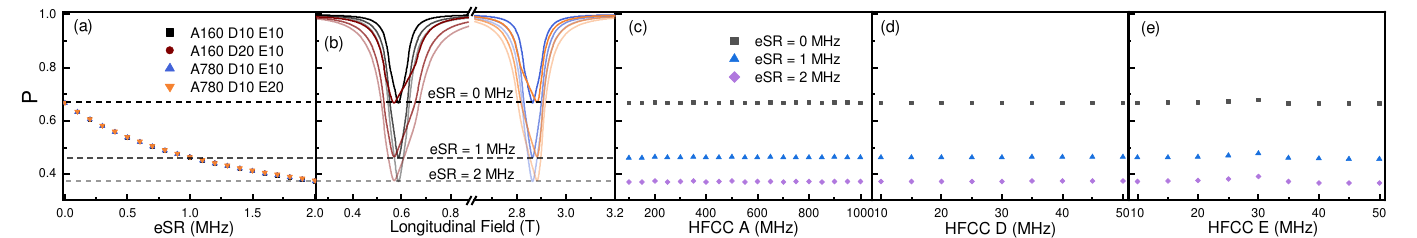}
\caption{\label{F SsingleALC}(a)The time-integral minimum polarisation versus eSR. (b) The corresponding time integral of the muon spin polarization curves with different eSR in 4 models. (c-e)The time-integral minimum polarisation versus HFCC A, D and E. (c) D = 10MHz and E = 10MHz. (d) A = 780MHz and E = 10MHz. (e) A = 780MHz and D = 10MHz. $P$ is polarisation and $P_{min}$ is the minimum polarisation (center of the ALC). }	
\end{figure*}

We used the Quantum software \cite{lord2006computer,arnold2014mantid} to model LF dependence of the muon's polarisation, $P(t)$, as a function of eSR, the isotropic (A) and anisotropic (D, E) hyperfine coupling constants (see \cite{nuccio2014muon} for a longer discussion) and for different combinations of ALCs with different muon-electron hyperfine constants. Quantum models the muon together with nearby spins such as electrons and nuclei, interacting via dipolar, hyperfine and quadrupolar interactions, taking into account of Zeeman interaction with both static and RF magnetic fields, and spin flips and site changes. It uses an effective Hamiltonian where the dynamic changes (e.g spin flips or site changes) are handled by a density matrix approach with exponential relaxation rates. Then different muonium sites are calculated independently, and combined to form a final field dependent signal to compare to the experimental data.

Shown in Figure\ref{F_ALC2}(a) is an example of the time dependence of the muon's polarisation, where it can be seen that the (fast) Rabi oscillations are present at all values of eSR, superimposed on the relaxation rate brought about by eSR. In most experiments, the Rabi oscillations are averaged out and not observed. A common way to plot the ALC resonance data taken on time differential (TD) muon spectrometers is to integrate the time dependent spectra. This is without detrimental effect to the underlying physics of the sample in most cases, but one must be aware of a few different effects. Firstly, the modelling and data analysis become easier, by virtue of only needing to worry about the model independent time integral (TI) ALC. The the overall statistics needed to be taken is lower because fitting in the time domain is model dependent and tends to weight slightly more at higher times, resulting in larger errors on parameters (in particular, the asymmetry). At pulsed muon sources, TI detector deadtime is less apparent than TD, allowing a greater acquisition rate. This makes beamtime more efficient, although large deadtime can still introduce non-linearities between the actual TI asymmetry and the measured one.

 Modelled TI asymmetry for some typical values of eSR is shown in Figure \ref{F_ALC2}(b).  The amplitude of the ALCs increases as eSR increases, due to the exponential polarisation loss related to the coupling of the muon's  and electron's spins. An experimentally convenient proxy therefore for measuring eSR is the amplitude of the ALC resonance (i.e minimum polarisation, $P_{min}$) \cite{schulz2011importance}. For those materials where the ALC position is not particularly temperature dependent, eSR can be extracted as a function of temperature by a single point measurement of $P_{min}$, with only full-field scans of the ALCs taken at a limited number of temperatures, saving a significant amount of experimental time \cite{schulz2011importance}. This technique has been demonstrated in our modelling; $P_{min}$ is shown as a function of eSR for various A, D and E's in Figure \ref{F SsingleALC}(a), with the corresponding ALC resonances in Figures \ref{F SsingleALC}(a) and (b). It is clear from the overlapping points in Figure \ref{F SsingleALC}(a), that the eSR extracted from $P_{min}$ is independent of A, D or E. Figure \ref{F SsingleALC} (d)-(f) demonstrates that $P_{min}$ shows minimal dependence on A, D or E over the parameter space relevant to the majority of muoniated radicals in OSCs. We can therefore extract eSR from single point measurements, once properly normalised (to the eSR = 0 assumption at low temperature \cite{schulz2011importance}), if  the ALC resonance doesn't move with temperature and there is only one dominant ALC resonance at the field chosen. It is worth noting that for the majority of experiments on solid state OSC, these conditions are not met.

\begin{figure}[!h]
\includegraphics[width=\linewidth]{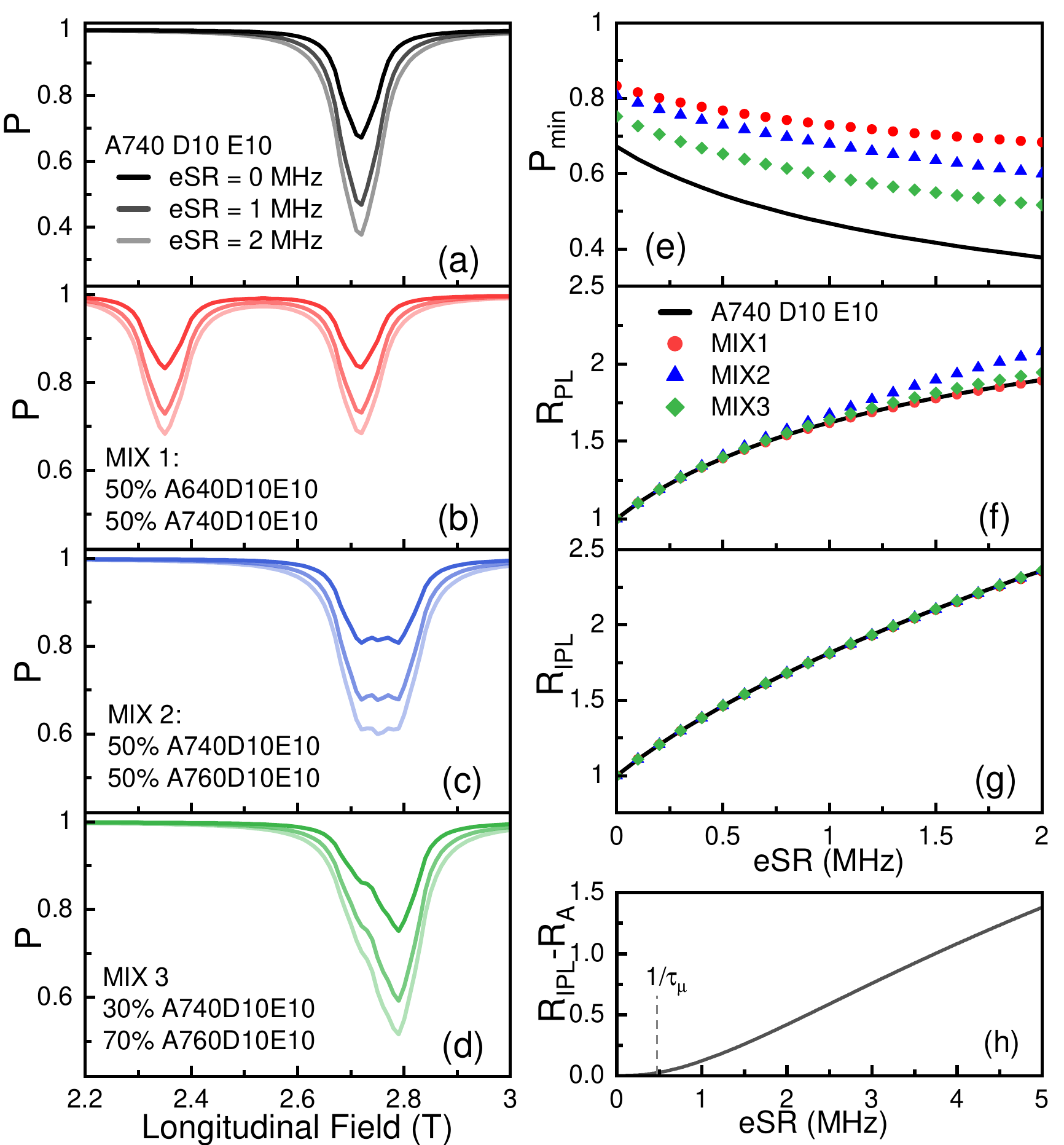}
\caption{(a)The single ALC resonance used for benchmark.(b)Two ALC resonances do not overlap. (c)(d)The same ALC resonances overlap with different composition.(e)-(g)The time-integral minimum polarisation, the minimum polarisation loss ratio and the integrated polarisation loss ratio versus eSR. (h) The difference between $R_{IPL}$ and a simple amplitude-only model, showing a crossover to line broadening dominating $R_{IPL}$ around $1/\tau_\mu$, which is roughly linear as eSR increases. See text.}
\label{f eSR2}
\end{figure}

Muon-electron hyperfine coupling constants can vary considerably between different muon aducts, even ones that are next to each other on the same molecule. For many molecular systems, several ALC resonances can overlap and form a wider and more complex signal; it is particularly a problem for those systems with large D and/or E, yielding merged and ill-defined ALC data.  We have calculated these effects for 1 or 2 ALC resonances using reasonable estimates of parameters, shown in Fig. \ref{f eSR2}(a)-(d) - a single ALC resonance used for benchmarking, and MIX1, MIX2 and MIX3 containing different combinations of ALC resonances. MIX1 is the combination of two ALC resonances of equal weight that are separated by a significant margin, but, still have their `tails' overlapping with each other in the field range between the two. MIX2 and MIX3 are the combination of two much closer ALC resonances, with the different between MIX2 and MIX3 being the ratio of the two amplitudes.

In Fig. \ref{f eSR2}(e), we can see the $P_{min}$ now has dependence on the specifics of the ALCs being measured, making extraction of eSR from it challenging. The majority of the difference between the benchmark and MIX1-3, is related to the signal being split between several ALCs. In typical experimental analysis, this would be normalised out. We define the maximum polarization loss ratio, $R_{PL}$, as $R_{PL} = \frac{PL_{eSR}}{PL_{0}}$, where $PL_{eSR} = 1 - P_{min}$ with a non-zero eSR and $PL_{0} = 1 - P_{min}$ with zero eSR. This is similar to the normalisation techniques used on real experimental data, and the outcome is shown in Figure \ref{f eSR2}(f). It can be seen that MIX1 (separated ALCs) now sits nicely onto the benchmark, but the other two do not. This unambiguously demonstrates that it is possible to improve efficiency of the temperature dependent measurements of the estimate the eSR by taking a single point on the ALC as a function of temperature, when the temperature dependence of A is small, there is only one ALC line dominating the signal and when eSR is relatively small.  

There another step we can take to improve matters. We define an integrated polarization loss ratio (meaning the ratio of integrated ALC resonances for finite eSR and 0 eSR) as
\begin{equation} 
 R_{IPL}=\left. {\sum\limits_{B_{min}}^{B_{max}}PL_{eSR}(B_i)\Delta B_i} \right/ {\sum\limits_{B_{min}}^{B_{max}}PL_{0}(B_j)\Delta B_j}  
\end{equation}
\noindent where $B_{min}$ and $B_{max}$ the minimum and maximum fields, chosen to minimise the systematic errors (see the Supplementary Material for how to choose them), $PL_{eSR}(B_i)$ and $PL_{0}(B_j)$ are the polarisation loss for the model with eSR at the longitudinal fields, $B_i$, and without eSR at the longitudinal fields, $B_j$. $\Delta B_i = \frac{1}{2}(B_{i+1} - B_{i-1})$. $R_{IPL}$ is shown in Fig. \ref{f eSR2}(g), where it is clear that the change in area underneath the ALC is independent of the technique hyperfine couplings (A,D,E), the number of ALCs present and whether they overlap or not. This therefore represents generalised tool, where the eSR falls onto a single universal law. It has the potential to simplify the analysis significantly. 

To understand the physics of how $R_{IPL}$ varies with eSR, we consider the following model. Without any eSR, only the radicals with electron spin up can contribute to the muon's relaxation. With sufficient relaxation that the electron is almost certain to be flipped in the muon lifetime (so initially spin down ones then contribute to the integral asymmetry), but still slow enough compared to the muon-electron hyperfine constants that it doesn’t effect the linewidth much, but does effect the overall amplitude. We would expect this to have a functional form of $R_A = 2-\exp{(-\tau_{\mu} eSR/2)}$. The difference between $R_{IPL}$ and $R_A$ is shown in Figure Fig. \ref{f eSR2}(h), where there is good agreement below about $1/\tau_\mu$, indicating that the primary effect at low eSR is amplitude. Now turning to the width increase as eSR gets larger, above $1/\tau_\mu$, when off the line centre the muon's polarisation oscillates around a value between 1/2 and 1. Flipping the electron to spin down results in the polarisation being ``frozen'' (strictly the longitudinal part is frozen parallel to B and the perpendicular parts are dephased by rapid precession) and if you flip it again the muon oscillation starts again from a phase of ``0 degrees'', scaled down by the frozen value. So the line should broaden, since the central asymmetry is only doubled but the sides are enhanced more. We note that for any individual muon site (fixed angle with respect to the principal axes of the muon-electron hyperfine tensor) the line is a simple Lorentzian $P=1/2 +  1/2A^2 / [(B-B_0)^2\gamma^2+A^2]$, if $A>1/\tau_{\mu}$, and it might give a simple answer dependent only on eSR, provided we also have $A>eSR$. To make further progress, a challenging integral is needed to be solved analytically, but in the supporting material we demonstrate the line-broadening appears to be roughly linear with eSR (within reasonable limits, below 5 MHz). For now, purely as an experimental tool, we report the best polynomial fit to this universal law as $eSR = \frac{1}{\tau_{\mu}}[1.982(R_{IPL}-1) + 1.028(R_{IPL} - 1)^2 ]$, likely to be a Taylor expansion to an exponential function.

In real experimental data, we usually cannot measure the full field range of the ALC resonances, since the `tails' of the ALC resonance have a smaller amplitude than the experimental errors for a reasonable measurement time and the extended field ranges needed would add significantly to the required beam time. There is no one-size fits all for advising on the choice of the field range to measure (or analyse the data),  as the widths of different ALC resonances are dependent on anisotropic HFCs, which are sample dependent. As a guide for experiments, how to deal with and define an effective field range (EFR) along with uncertainties introduced by having an too small an EFR is discussed in the Supplementary Material.

\begin{figure}[!h]
	\includegraphics[width=\linewidth]{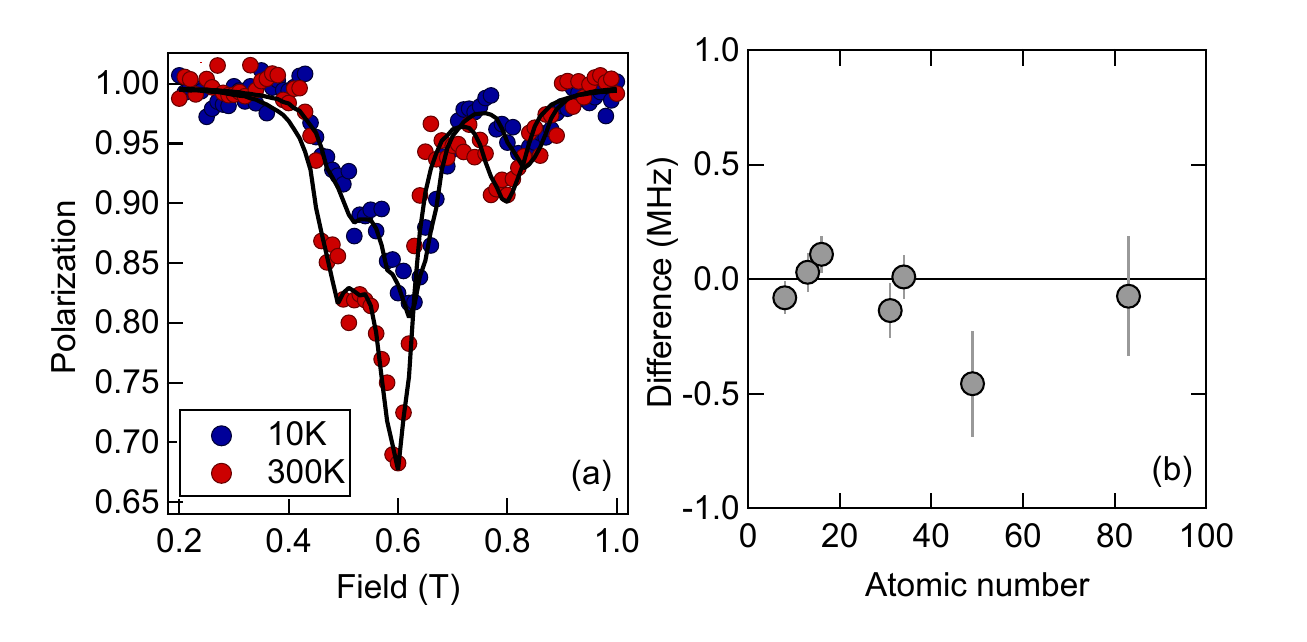}
	\caption{ (a) A relatively complex set of overlapping ALCs in triethylsilylethynyl anthradithiophene. Similar complexity and the same trends are present in all of the data in \cite{nuccio2013importance} (b) The difference between the new (this paper) and old (\cite{nuccio2013importance}) method of extracting eSR, plotted against the atomic number of the heaviest element in the molecule. There is clearly good agreement between the two methods. Seven samples in two series are shown: tris(8-hydroxyquinolinato) Aluminium/Gallium/Indium/Bismuth and triethylsilylethynyl anthradifuran/anthradithiophene/anthradiselenophene. EFR was chosen to be the largest range available given the data, and varied from about 2.5 up to 10, depending on the dataset. The point at an atomic number of 50 appears to have a difference of 0.5MHz, although the uncertainties are large (mainly due to difficulty in modelling) and is still within 2$\sigma$.}
	\label{f5}
\end{figure}

To test our method on real data, we have analysed previously published work \cite{schulz2011importance}, which showed that eSR at room temperature (on the MHz scale) increased linearly with atomic number of the central atom of two metal-organic complexes, obtained via lineshape modelling of complex ALC spectra. One example ALC spectra is shown in Figure \ref{f5}a. Figure \ref{f5}b shows the difference between the eSR extracted using the previous modelling method and the method in this paper, for seven OSC containing metals of different atomic numbers. The eSR is plotted against the  atomic number of the heaviest element on the molecule - the metal. There is strong agreement between the old and new methods.

There are a number of discussion points regarding this method. Whilst it is possible to handle a wider selection of data, in particular involving overlapping ALCs, it does only give you the average eSR of those overlapping ALCs (but if there are separated ALCs, then it's the average within each of those separated groups).  Given for each adduct the electron is potentially spatially restricted to different and specific parts of the molecule, this might result in different eSRs that are then averaged. This is particularly true for those molecules with heavy elements and a complex structure, as different adducts may result in different electron wavefunction overlap on the heavy element, different SOC, and thus different eSR. Modelling/fitting the ALC lineshapes and field dependent time series independently of one another may differentiate between the different eSR's and so should be the preference when undertaking analysis, but it is not always possible. Moreover, another source of relaxation such as impurities or structural changes modifying muonium sites or bonding probabilities/rates may significantly effect results.

There may also be issues with imperfect background subtraction; careful Ag backgrounds taken at all temperatures would help alleviate this problem, although these measurements are rarely perfect. If the background at different temperatures (or between samples) is identical in the raw data, then the effect will cancel out - this may be true for titanium cells for liquid experiments, but is unlikely the case for solid-state experiments in Ag packets. If an ALC falls on the repolarisation curve, most relevant in systems with at least two sites with one muon-electron isotropic HFI interaction being large and the other small, then the repolarisation curve can be treated as a background if it isn't possible to model it directly.

Nonetheless, this new analysis technique does have some significant advantages. In the case where complex ALC signals are present (overlapping ALCs) with similar eSRs then this gives us an additional tool in our analysis arsenal. In some cases we cannot ascertain whether a given ALC spectra is from an individual ALC resonance or several ALC resonances merged. Direct modelling in this case is challenging; single-point analysis of the temperature dependence may be impossible. This technique is particularly relevant in certain polymers, where significant structural disorder leads to unresolvable ALCs \cite{he2015muon}.  By comparing the results from $R_{PL}$ with the $R_{IPL}$, we may be able to determine whether a particular experimental signal is a single ALC resonance or multiple overlapping ALC resonances, which is helpful to build a correct model. We would also be able to compare results from $R_{IPL}$ with modelling to act as a benchmarking exercise; agreement between these two independent methods of extracting eSR would strengthen the conclusions that one can make from the modelling, and vice versa.

 LZ acknowledges financial support from the Chinese Scholarship Council.

\bibliography{ESRExtract_D1}
\end{document}


\maketitle

\section{Effective Field Range(EFR)}

\begin{figure}[!h]
\includegraphics[width=\linewidth]{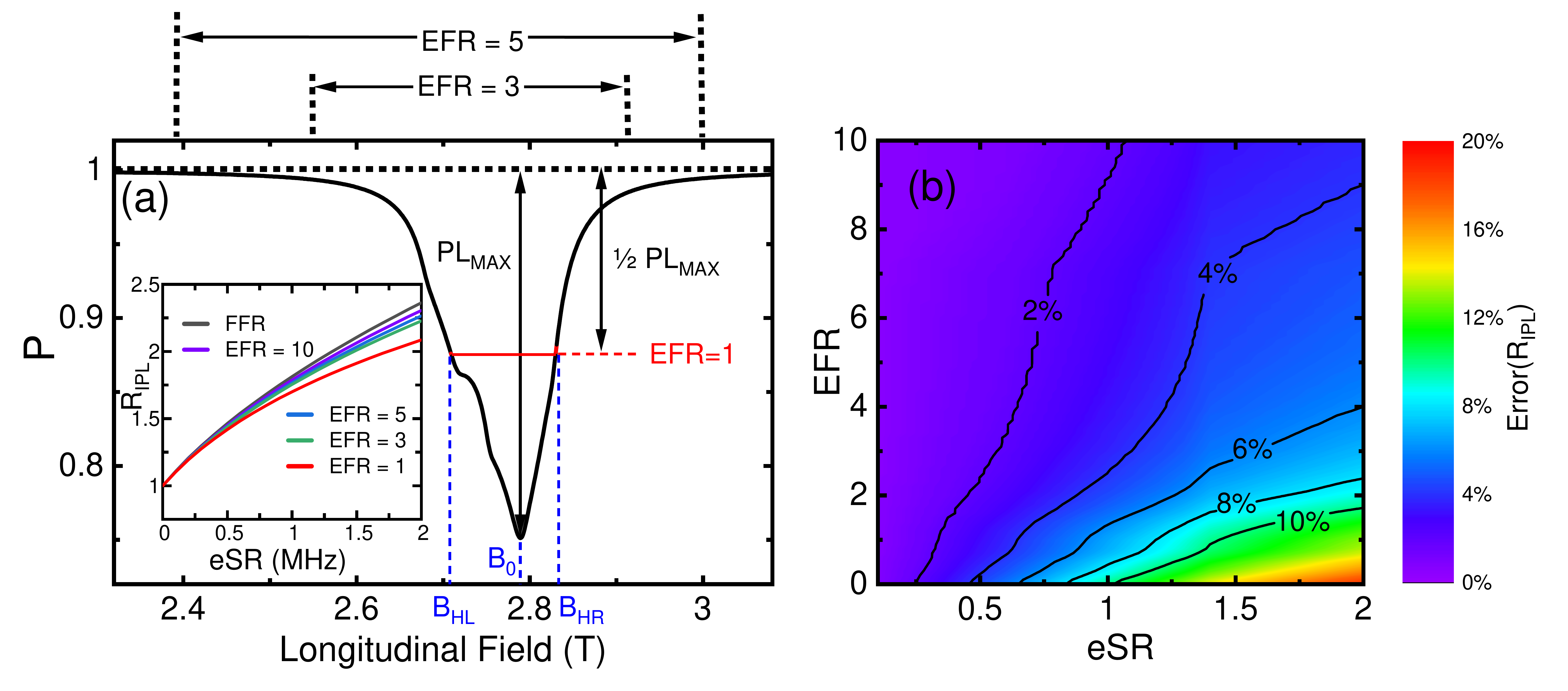}
\caption{(a)The effective field range of the overlapped ALC resonances MIX3. Insert plot: The integrated polarisation loss ratio versus eSR under different effective field range. (b)The integrated polarisation loss ratio error versus eSR and EFR.}
\label{f4}
\end{figure}

In real experimental data, we usually cannot measure the full field range(FFR) of the ALC resonances, since the `tails' of the ALC resonance have a smaller amplitude than the experimental errors for a reasonable measurement time and the extended field ranges needed would add significantly to the required beam time. There is no one-size-fits-all for advising on the choice of the field range to measure (or analysing the data),  as the widths of different ALC resonances (in the solid state) are dependent on anisotropic HFCs, which are sample dependent. We introduce an effective field range (EFR) defined by high and low fields, respectively $B_{HL}$ and $B_{HR}$. We define EFR = 1 when the polarization loss at the field point $B_{HL}$ and $B_{HR}$ is one half of the maximum polarization loss, or, EFR = 1 at the full width half maximum. This is demonstrated in Fig.\ref{f4}(a) as the red line.

For an arbitrary field range from $B_{L}\leq B_{0}$ to $B_{R}\geq B_{0}$), where $B_{0}$ is the peak ALC field,
$EFR = 1/2((B_{0}-B_{L})/(B_{0}-B_{HL})+(B_{R}-B_{0})/(B_{HR}-B_{0})).$
The inset to Fig.\ref{f4}(a) shows the effective field range of some overlapped ALC resonances (using the same parameters as in MIX3 in the main manuscript), and $R_{IPL}$ (also defined in the main manuscript) as a function of eSR for different EFRs. It is clear that the accuracy of $R_{IPL}$ as a method to extract eSR is dependent on the value of EFR. To quantitatively understand the limits of $R_{IPL}$, we define the error as
$Error = (R_{IPL}(FFR)-R_{IPL}(EFR))/R_{IPL}(FFR)$, which is shown as a percentage error in Fig.\ref{f4}(b). Since we are able to model the ALC lines, and access the tails of the ALCs, this is a good way to understand the limits of experiments and our analysis of real experimental data. This error can be used as a guide for performing experiments. It is clear that over a large amount of parameter space relevant for OSC (i.e eSR $\leq$ 1), the error on the measurements of eSR are only a few percent. This is a systematic error, and likely lower than any error associated with statistical uncertainty in the measurements for most experiments over most of the range of parameters shown in Fig.\ref{f4}(b). On a typical experiment where eSR is relatively low, it is therefore prudent to measure to an EFR$>$2, but for large eSR one needs to measure to much wider field ranges, as a result of the width and intensity increases of the data when eSR is large.

\begin{figure}[!h]
	\includegraphics[width=\linewidth]{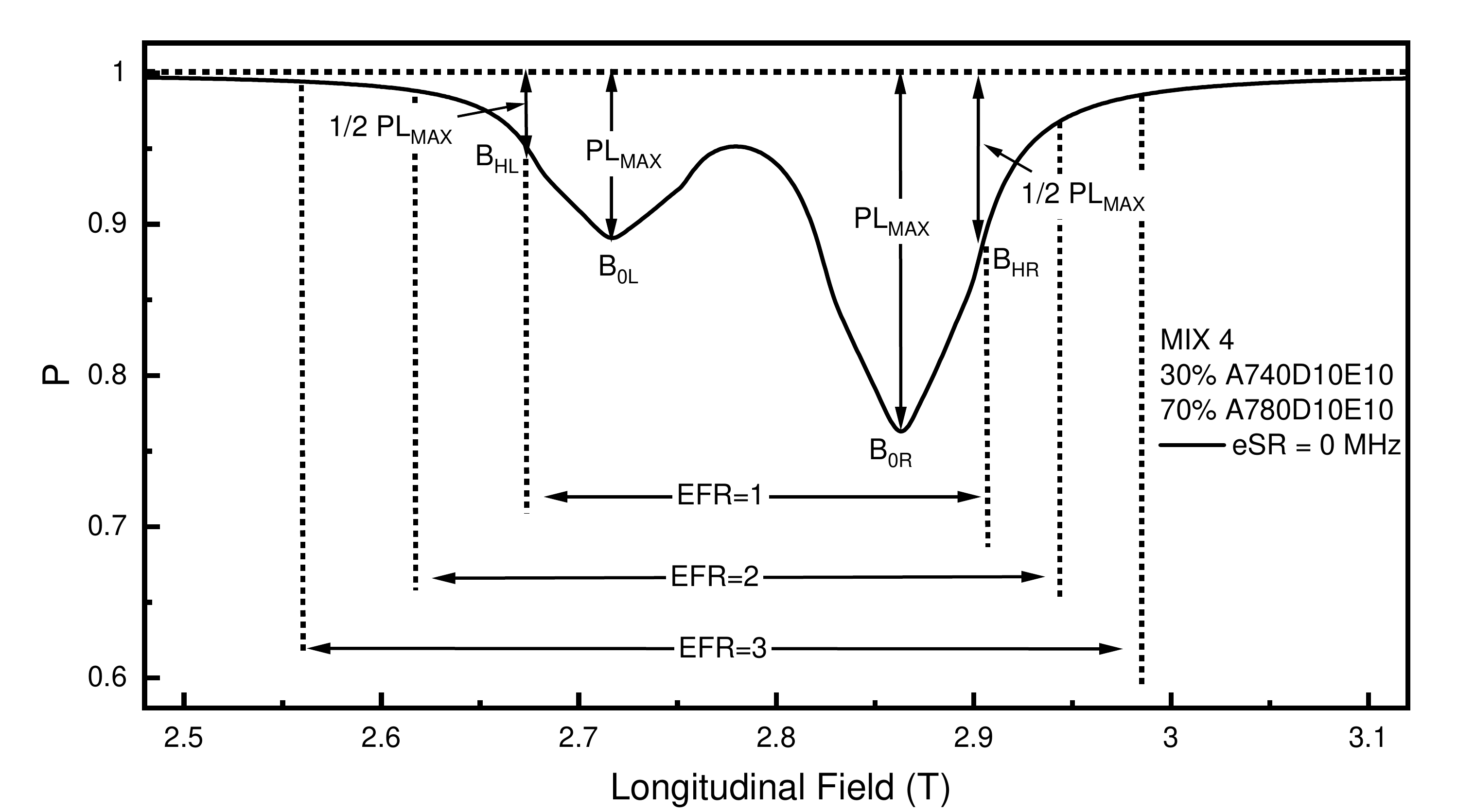}
	\caption{The effective field range of the overlapped ALC resonances with two peaks}
	\label{smf2}
\end{figure}
 There may be some ALC resonances overlapping but showing separate peaks in the signal. For this situation, one must set the field range between different peaks as a shared field range.
$B_{HL}$ is the field point where the polarization loss is one half of the maximum polarization loss of the left peak on the left side, and $B_{HR}$ is the for the right side. $B_{0L}$ is the field point where the minimum polarization is present for the left peak and $B_{0R}$ is for the right peak. This is demonstrated in Figure \ref{smf2}.  We still set the field range from $B_{HL}$ to $B_{HR}$ as EFR = 1, and for a selected field range from $B_{L}$ to $B_{R}$ where $B_{L}\leq B_{0L}$ and $B_{R}\geq B_{0R}$, such that $EFR = 1/2((B_{0L}-B_{L})/(B_{0L}-B_{HL})+(B_{R}-B_{0R})/(B_{HR}-B_{0R})).$

\section{Details of $R_{IPL}$  calculation}

We utilized the Quantum software within the platform Mantid for the simulation\cite{lord2006computer,arnold2014mantid}. Once Mantid is installed, go to the File option and open the Script Repository and then download the files in the section Muon/Quantum, including the subdirectories. One can run the scripts independently in Python, without the need to use the Mantid interface. Below is a typical example of the Python code used for generating our model from Quantum:
\begin{minted}{Python}
import numpy as np
from quantumtabletools import RunModelledSystem

Rlist=[]
for A in range(400,401,5): # the HFCC A
    for D in range(20,21,5): # the HFCC D
        for E in range(10,11,5): # the HFCC E
            for e in range(0,51,10): # eSR
                eSR=e/10 # eSR
                # setup of Quantum table
                Tab={}
                Tab['spins']=['Mu','e']
                Tab['dynamic']=[1]
                Tab['a(@0,Mu)']=[A,D,E]
                Tab['relax(@0,e)']=[eSR]
                Tab['lfuniform']=[25]
                Tab['loop0par']='bmag'
                Tab['loop0range']=[0,2.6,21]
                Tab['measure']=['integral']
                #run Quantum simulation and read back output
                xaxes,ydata,edata = RunModelledSystem(Tab)
                Rlist.append([A, D, E, eSR, ydata[0].tolist()])

#save output
OutputfileName='Mue_loopADEeSR_Model2.txt'
f=open(OutputfileName,'w')
for i in range(len(Rlist)):
    s=map(lambda x: "{:.5f}".format(x),Rlist[i][4])
    print (Rlist[i][0],Rlist[i][1],Rlist[i][2],Rlist[i][3],*s, file=f)
f.close()

\end{minted}
You will obtain the LF dependence of the muon’s polarisation, $P(B)$, as a function of eSR, the isotropic (A) and anisotropic (D, E) hyperfine coupling constants.

To calculate the integrated polarization loss ratio, as described by Eq.(1) in the article, it is necessary to determine the full field range of the ALC resonance from $B_{min}$ to $B_{max}$, the minimum and maximum fields chosen to minimise the systematic errors. Specifically, reducing $B_{min}$ (unless the ALC resonance falls within the repolarization field range) or increasing $B_{max}$ would not alter the integrated polarization loss with 0 eSR beyond a reasonable level of precision. Once $B_{min}$ and $B_{max}$ are determined, the integrated polarization loss ratio $R_{IPL}$ can be easily calculated according to Eq.(1) in the article.

\section{Details of Alq3 data analysis}
\begin{figure}[!h]
	\includegraphics[width=\linewidth]{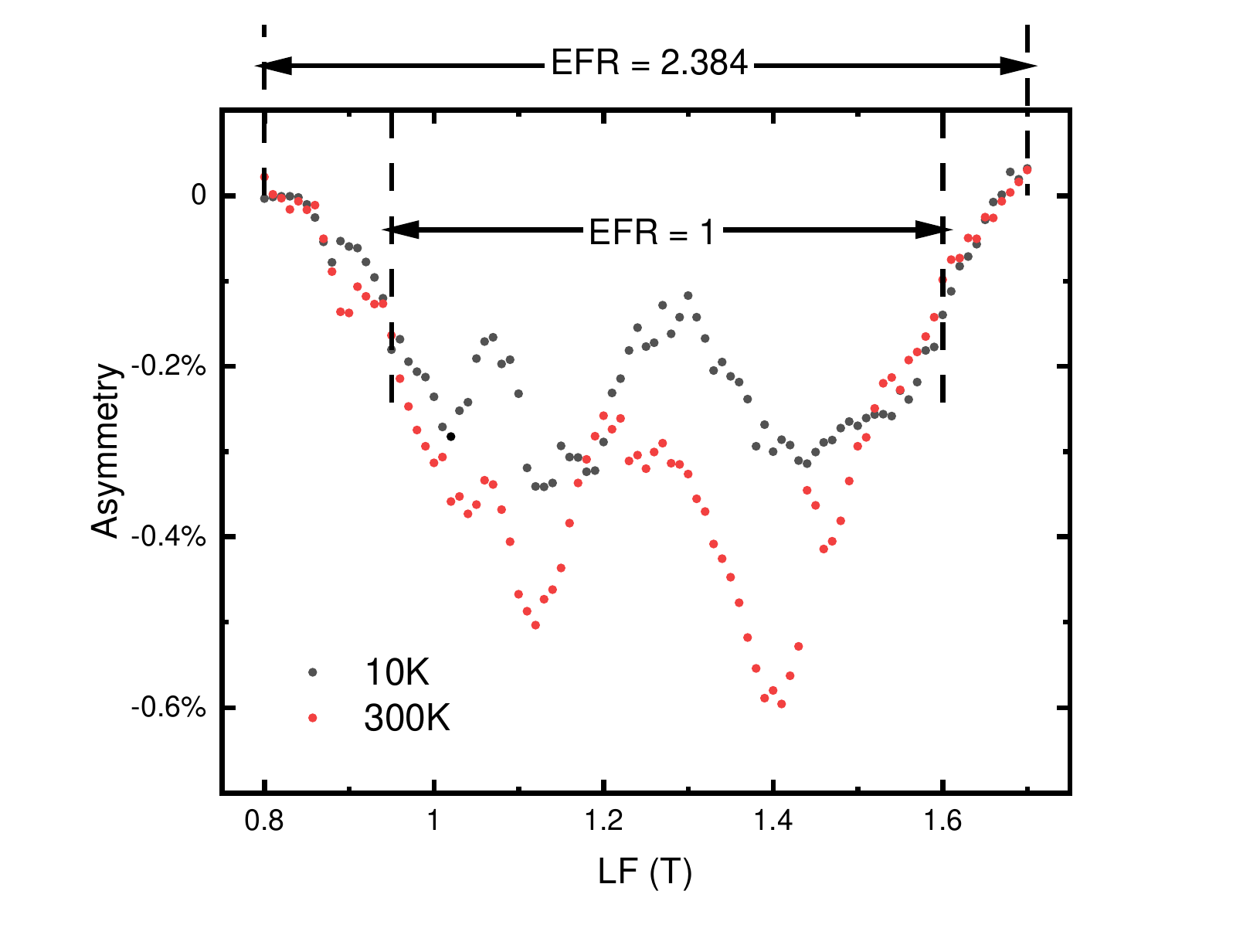}
	\caption{The asymmetry of Alq3 after background subtraction.}
	\label{SMF5}
\end{figure}
 Here, we utilize muon experimental data of Alq3, comprising tris(8-hydroxy-quinolinato) Aluminium, as an illustrative example for analysis and discussion. In muon experiments, the asymmetry of the spatial distribution of the positron provides a direct measurement of the spin polarization of an ensemble of muons in the sample environment at any time when it decays. The measured asymmetry $A(t)$ at $t=0$ is essentially equivalent to the asymmetry factor $a$ when the initial polarization of muon is 100\% antiparallel to its momentum. Thus, we have $A(t)=aP_{\mu} (t)$. Consequently, the asymmetry loss is directly proportional to the polarization loss, and the ratio between them eliminates the scale factor. This property significantly simplifies the extraction of the eSR from the ALC resonances.

Fig.\ref{SMF5} presents the asymmetry of Alq3 after subtracting the background. In our analysis, we assume the eSR at 10K is 0.02MHz or lower. The experiments were run within a limited field range, specifically from 0.8T to 1.7T. $B_L$ = 0.8T, $B_{R}$ = 1.7T. The values of various magnetic fields are as follows: $B_{0L}$= 1.02T, $B_{0R}$=1.44T, $B_{HL}$=0.95T, $B_{HR}$=1.6T. By using the equation  $EFR = 1/2((B_{0L}-B_{L})/(B_{0L}-B_{HL})+(B_{R}-B_{0R})/(B_{HR}-B_{0R}))$, the effective field range (EFR) of Alq3 is 2.384.

Through a integral calculation of the asymmetry loss, we obtain that the integrated asymmetry loss of Alq3 at 10K is -0.16542, while at 300K it is -0.24237. So the integrated polarization loss ratio $R_{IPL}$ is 1.465, corresponding to an eSR value of 0.52MHz. This result is in strong agreement with the original method that yielded an eSR value of 0.5MHz \cite{nuccio2013importance}.

\section{ALC width as a function of eSR}
\begin{figure}[!h]
	\includegraphics[width=\linewidth]{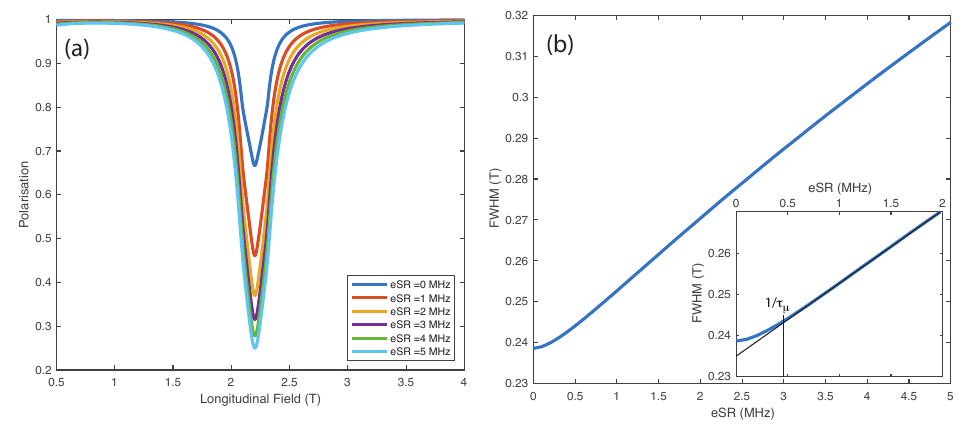}
	\caption{(a) Representative ALCs calculated for A=600 MHz and D=E=25 MHz for several different eSRs. (b) The dependence of the full width half maxima (FWHM) on eSR for ALCs calculated with A=600 MHz and D=E=25 MHz. This is nearly linear above approximately eSR=1/$\tau_\mu$. Inset: zoom in of the data in the main figure. The blue line is the modelled data, the black lines are linear guides to the eye.}
	\label{ALCeSR}
\end{figure}

We have performed calculations using the Quantum software to investigate how the width of an ALC varies with eSR. Figure \ref{ALCeSR}(a) shows some representative ALC curves, where it is clear that the amplitude increases significantly at low values of eSR, and continues to increase (but at a lower rate) as the eSR increases further. Figure \ref{ALCeSR}(b) shows how the full width half maximum of the ALCs vary with eSR, where it is clear that there is a nearly linear dependence above approximately $1/\tau_\mu$.

\bibliography{SM}
\bibliographystyle{unsrt}